\documentclass[12pt,oneside]{article}
\usepackage{graphicx}
\def\be{\begin{equation}}
\def\ee{\end{equation}}
\def\ba{\begin{array}}
\def\ea{\end{array}}
\def\bea{\begin{eqnarray}}
\def\eea{\end{eqnarray}}
\def\<{\langle}
\def\>{\rangle}
\def\~{\tilde}
\def\s{\sigma}

\newcommand{\av}[1]{\mbox{Av}\left(#1\right)}

\begin{document}

\title{Energy-Decreasing Dynamics\\
in Mean-Field Spin Models}

\date{}
\maketitle

\begin{center}
{\large L. Bussolari, P. Contucci, M. Degli Esposti, C. Giardin\`a}\\
\vspace{0.3cm}
{\small\it
Dipartimento di Matematica dell' Universit\`a di Bologna,\\
Piazza di Porta S. Donato 5, 40127 Bologna, Italy \\
}
\end{center}
\vspace{2.0cm}
\begin{abstract}
We perform a statistical analysis of deterministic energy-decreasing
algorithms on mean-field spin models with complex energy landscape like
the Sine model and the Sherrington Kirkpatrick model. We specifically
address
the following question: in the search of low energy configurations
is it convenient (and in which sense) a quick decrease along the gradient
(greedy dynamics) or a slow decrease close to the level curves (reluctant
dynamics)?
Average time and wideness of the attraction basins are introduced
for each algorithm together with an interpolation among the two
and experimental results are presented for different system sizes.
We found that while the reluctant algorithm performs better for a fixed
number 
of trials, the two algorithms become basically equivalent for a given
elapsed 
time due to the fact that the greedy has a shorter relaxation time which
scales
linearly with the system size compared to a quadratic dependence for the
reluctant.
\end{abstract}

\section{Introduction}

The problem of finding the ground state of a frustrated
spin model having a complex energy landscape is, in general,
an NP-complete problem: the running time of exact algorithms
increases at least exponentially with system size.
There are, however, several new ground state techniques
devised for specific examples which are able to
calculate exact ground state in a polynomial time
using elementary algorithms in combinatorial optimization,
in particular network flows \cite{R1,HR}.
This opened the route to the numerical study of very large
system sizes for different problems, like
spin-glasses \cite{R2}, random field Ising model \cite{O},
solid-on-solid model with a disordered
substrate \cite{BHMR}, superconducting flux line lattices
\cite{GPK}, and many others.
In the general case,
where such particular algorithms are not known, one is forced to
use approximated methods. These consists
in some kind of dynamic in the space of spin
configurations which explores different states looking for
the lowest energy value. The simplest choice is to
consider some kind of Monte Carlo simulation at zero
temperature (deep quench) which,
starting from a random configuration, follows a random walk
of decreasing energy till one ends up on a local energy minimum.
One then repeats 
this procedure a number of time as large as possible
and take as better estimate of ground state
the lowest energy found. Many variants and
improvements have been proposed and among them
the Simulated Annealing \cite{KGV},
which slowly cool the system from high temperature
to zero temperature and Parallel Tempering, which uses
several temperatures in parallel\cite{MP,HN}.
Which algorithm is most suitable depends on the
nature of the problem; for a recent paper, where the
performances of these different Monte Carlo simulation
techniques are compared, see ref. \cite{MKH}.
 
Monte Carlo dynamics, of one type or another, is
{\em stochastic}, i.e. for a given (random)
initial configuration the trajectory is a
random process. On the other hand, to find
low-temperature states, one may also consider
{\em deterministic} dynamics, which uniquely
associates to a (random) initial spin configuration
a final state according to some evolution rule.

A simple question which naturally arises is the
following: what kind of deterministic dynamic
is most effective in finding the configurations
of smallest energies? While stochastic dynamic have
been widely studied in literature, much less is known
on statistical properties of deterministic dynamic.

In this paper we focus our attention on two of
them: {\it greedy} and {\it reluctant}:
both of them follows a one-spin-flip decreasing
energy trajectory, the difference being that while
in greedy dynamics the energy decreases of the largest
possible amount, the reluctant algorithm makes
moves corresponding to the smallest possible energy decrease.
Some of the properties of these
two minimization algorithms were studied
in \cite{P}. In this paper
we push further the analysis addressing the following questions:
\begin{itemize}
\item 
For a given number of initial spin configurations
which of the two dynamics is more efficient?
Which one has the largest basin of attraction?
\item
For a given elapsed time which one is able to reach
the lowest energy states?
\end{itemize}

\noindent
In order to answer these questions we considered
two different models. For the Sine model (Section 3), where it is
available an analytical knowledge of ground
state for particular values of system sizes,
we focus our attention on the capability of the
two algorithms in detecting this ground
state. For the Sherrington-Kirkpatrick model we present numerical
results in Section 4, where the lowest energy found
is studied with different parameters in the
simulations. The outcome of the analysis
is that while for a fixed number of initial
spin configurations the reluctant dynamics
works better as it was found in ref. \cite{P}
(there is a higher probability to find low energy configurations),
when the elapsed running time is fixed the two algorithms gives
basically the same results (the time used for a single run
grows linearly for the greedy algorithm and quadratically for
the reluctant).
A final test is also performed in a stochastic
convex combination of the two algorithms: at each step the
motion is greedy with probability $P$ and reluctant
with probability $1-P$. It is found that for large $N$
and for fixed running times a substantial improvement
is obtained with a $P=0.1$.

\section{Greedy and Reluctant Dynamics}

We consider models defined by the Hamiltonian
\be
H(J,\s) = - \frac{1}{2}\sum_{i,j=1}^N J_{ij}\s_i\s_j
\ee
where $\s_i=\pm 1$ for $i=1,\dots,N$ are Ising spin variables
and $J_{ij}$ is an $N\times N$ symmetric matrix which
specifies interaction between them.

The {\em greedy} and {\em reluctant} dynamics work as follow.
The initial spin configuration at time $t = 0$
is chosen at random with uniform probability.
Then the evolution rule is:

\begin{enumerate}

\item 
Let $\s(t) = (\s_1(t),\s_2(t),\ldots,\s_N(t))$ be the
spin configuration at time $t$.

\item
Calculate the spectrum of energy change obtained by
flipping the spin in position $i$, for $i=1,\ldots,N$:
\be
\Delta E_i = \s_i(t)\sum_{j \neq i} J_{ij} \s_j(t)
\ee

\item
Select the site $i^{\star}$ associated with the lowest
(resp. highest) of the {\em negative} energy change for
the greedy (resp. reluctant) dynamic.
\be
i^{\star}_{greed} = \left\{ i \in \{1,\ldots,N\} :
\Delta E_{i^{\star}} =
\min_{i \in \{1,\ldots,N\}} \{ \Delta E_i < 0 \} \right\}
\ee 
\be
i^{\star}_{reluc} = \left\{ i \in \{1,\ldots,N\} :
\Delta E_{i^{\star}} =
\max_{i \in \{1,\ldots,N\}} \{ \Delta E_i < 0 \} \right\}
\ee 

\item
Flip the spin on site $i^{\star}$:
\be
\s_i(t+1) = \left\{
\begin{array}{ll} 
-\s_i(t), & \mbox{if $i=i^{\star}$}, \\
\s_i(t), & \mbox{if $i \neq i^{\star}$},
\end{array}
\right.
\ee

\end{enumerate}

\noindent
Both the dynamics follows an energy descent trajectory
till they arrive to a $1$-spin-flip stable configuration,
i.e. a configuration whose energy can not be decreased
by a single spin-flip. These represent local
minima in energy landscape at zero temperature with
respect to a $1$-spin-flip dynamic. They are also
solutions of the mean field TAP equations at zero
temperature \cite{TAP}:
\be
\s_i = \mbox{sign} \left(\sum_{j \neq i} J_{ij} \s_j\right)
\ee

\section{Results for the Sine model}

The first model we study is a mean-field system,
having a very high degree of frustration even tough
the bonds between spins are non random. It has been
introduced by Marinari, Parisi and Ritort in ref.
\cite{MPR1,MPR2}.
The couplings are given by the orthogonal
matrix associated with the discrete Fourier transform:
\be
J_{ij} = \frac{2}{\sqrt{2N+1}}\sin\left(\frac{2\pi ij}{2N+1}\right)
\ee
The ground state of the model is not known for general
values of the system size. However, as already noted in
\cite{MPR1,MPR2},
for special values of $N$ the
ground state can be explicitly constructed
using number theory.
Indeed,
for $N$ odd such that $p = 2N+1$ is prime of the form $4m+3$,
where
$m$ is
an integer, let $\s^L$ be the state given
by the  sequence of Legendre symbols, i.e.
\be
\s^L_i = \left (\frac{i}{p}\right ) = \left\{
\begin{array}{ll} 
+1, & \mbox{if $i=k^2 (\bmod p)$}, \\
-1, & \mbox{if $i \neq k^2 (\bmod p)$},
\end{array}
\right.
\ee
with $k=1,2,\ldots,p-1$. Then, it's easy to verify (see \cite{DGGI})
that
\be 
\label{gs}
H(\s^L)=-{N\over 2}
\ee
which is, of course, the lowest value that energy can take
due to the orthogonality of interaction matrix.

The explicit knowledge of ground state for selected
$N$ is a valuable bonus, since it allows
a complete control of dynamical and statistical
properties of the algorithms for quite large system
sizes. In this Section we restrict our analysis
to such $N$ values for which we have an exact
expression of ground state configuration as
Legendre symbols.
The natural unit of time is the
``spin-flip'' time, i.e. the cycle during which the dynamic explores all
the internal fields in such a way to decide which spin
to flip. In this unit the time $t$ of a realization
of the dynamic for a given initial condition is obtained
by counting the number of ``spin flip'' necessary to reach
a metastable configuration.

We run greedy and reluctant dynamics for a large
number $M$ of initial conditions, keeping track
of the number of times $n_{GS}$ we found the ground
state as final configuration. We also measured the time
of each realization $t_i$, $i = 1,\ldots, M$.
The number of trials $M$ is an increasing function
of the system size. For small sizes we stopped
when we found the ground state $10000$ times.
For the largest size $(N =  69)$ we used up to $10^9$
initial configurations, so that the ground state
has been found at least $100$ times.
We computed the following three quantities:

\begin{enumerate}
\item the average relaxation time of the dynamic
\be
\label{time}
\tau = \frac{1}{M}\sum_{i=1}^{M} t_i
\ee
\item the estimated probability to find the ground state
\be
\label{prob}
p_{GS} = \frac{n_{GS}}{M}
\ee
\item the average time to find the ground state
\be
\label{av_time_gs}
T_{GS} = \frac{1}{n_{GS}}\sum_{i=1}^{M} t_i
\ee
\end{enumerate}

\noindent which are obviously related by $T_{GS} = \tau /p_{GS}$.

\begin{figure}[h]
\begin{center}
\includegraphics[width=10.cm,angle=-90]{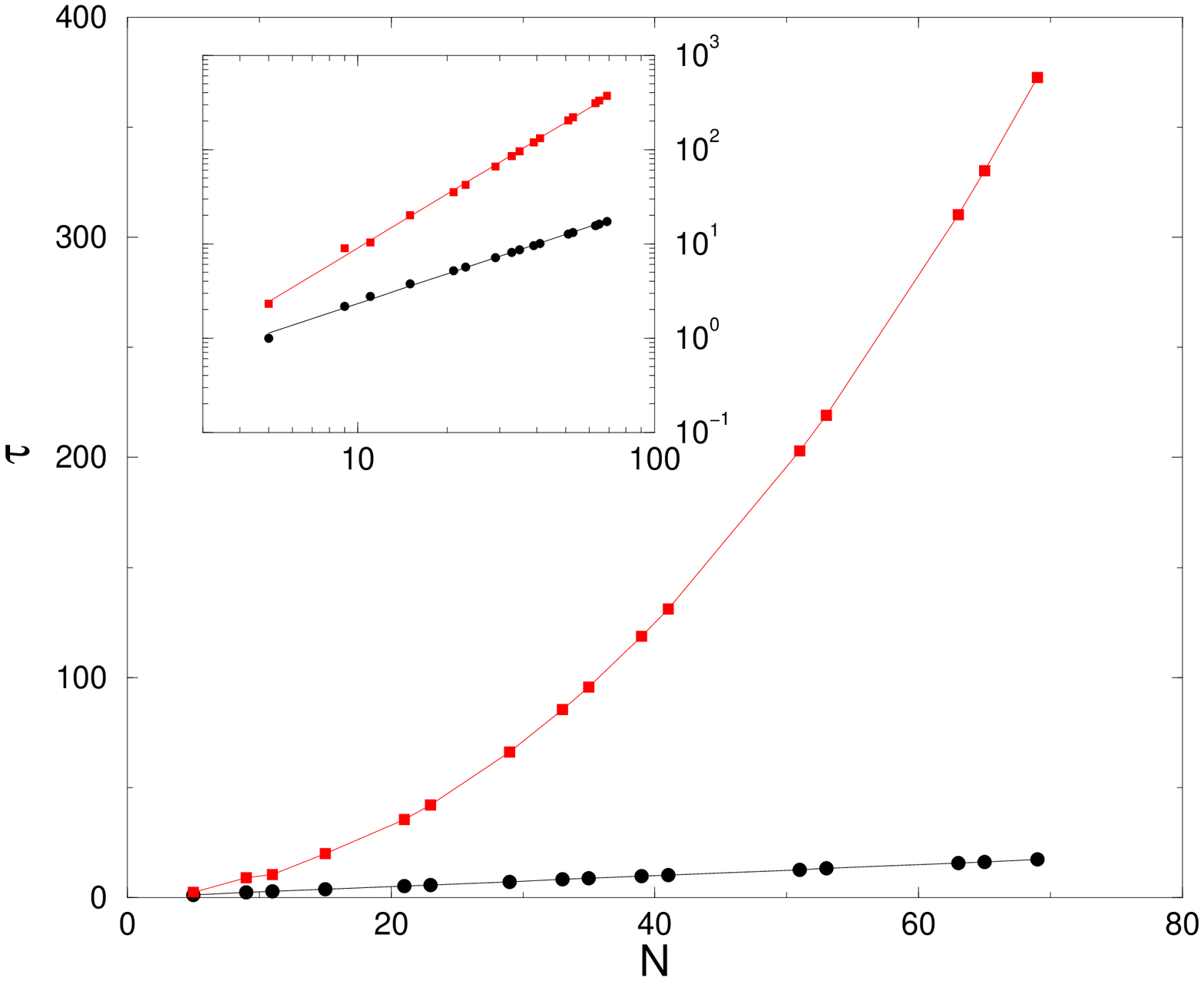}
\caption{
{\small The average time to reach a metastable configuration
for greedy (circle) and reluctant (squares) dynamics for the
Sine model. The inset
show the data in log-log scale. The continuous lines are the
numerical fits: $\tau_{gre}(N) \sim 0.25 N $ and
$\tau_{rel}(N) \sim 0.10N^{1.90}$}
}
\label{seno-intrinsic}
\end{center}
\end{figure}

\noindent
In Fig. (\ref{seno-intrinsic}) we plot the average
time of the dynamic to reach a metastable configuration.
As one could expect the greedy dynamic is much faster,
since it follows the most rapid path to decrease
energy.
The greedy average time is linear with the system size, while
the reluctant dynamic has a characteristic time which
increases as $N^{\alpha}$ with $\alpha\sim 1.90$.

\begin{figure}[h!]
\begin{center}
\includegraphics[width=10.cm,angle=-90]{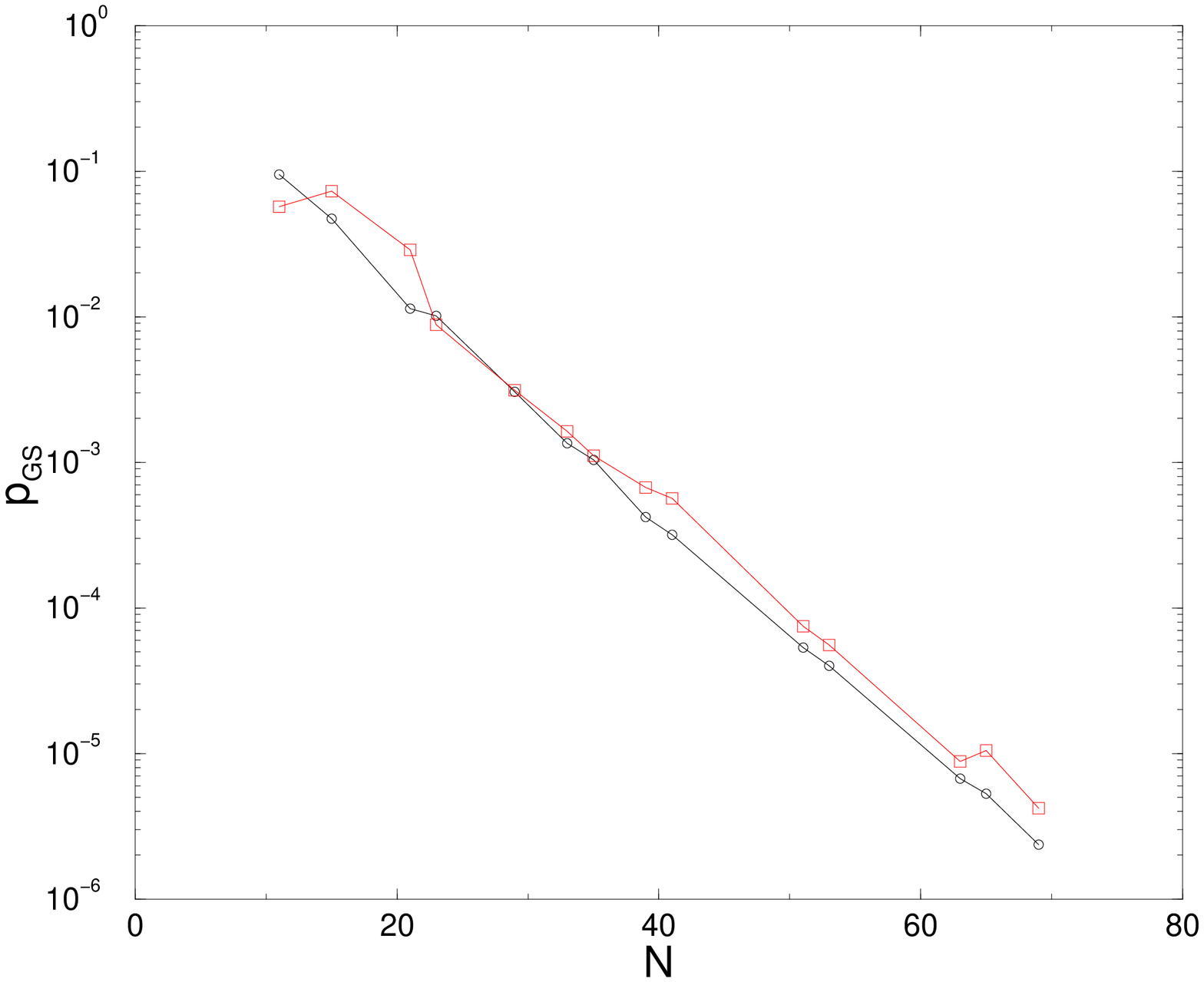}
\caption{
{\small Probability of finding the ground state for
greedy (circle) and reluctant (squares) dynamics for the
Sine model.}
}
\label{seno-prob}
\end{center}
\end{figure}

\noindent
In Fig. (\ref{seno-prob}) we compare the probabilities
of finding ground state for the two algorithms.
These have been estimated empirically using
formula (\ref{prob}). We always used a number
of trials $M$ large enough to ensure the robustness of the
statistical properties i.e. we increased $M$ until
the estimated probabilities relaxed to an asymptotic
value with negligible fluctuations.
We can see that, apart from finite size
effects for  very small system sizes, both algorithms have an
exponentially decreasing probability of finding
the ground state. Nevertheless the reluctant
probability is a little bit larger which says
that, for a given number of initial conditions,
reluctant algorithm is more efficient in
finding the ground state,
i.e. it has a larger basin of attraction.
This result agrees with the one in ref. \cite{P},
which using both algorithms with the same
number of initial conditions obtained a better
estimate of asymptotic value of energy
ground state for SK model in the case
of reluctant dynamic.

\begin{figure}[h]
\begin{center}
\includegraphics[width=10.cm,angle=-90]{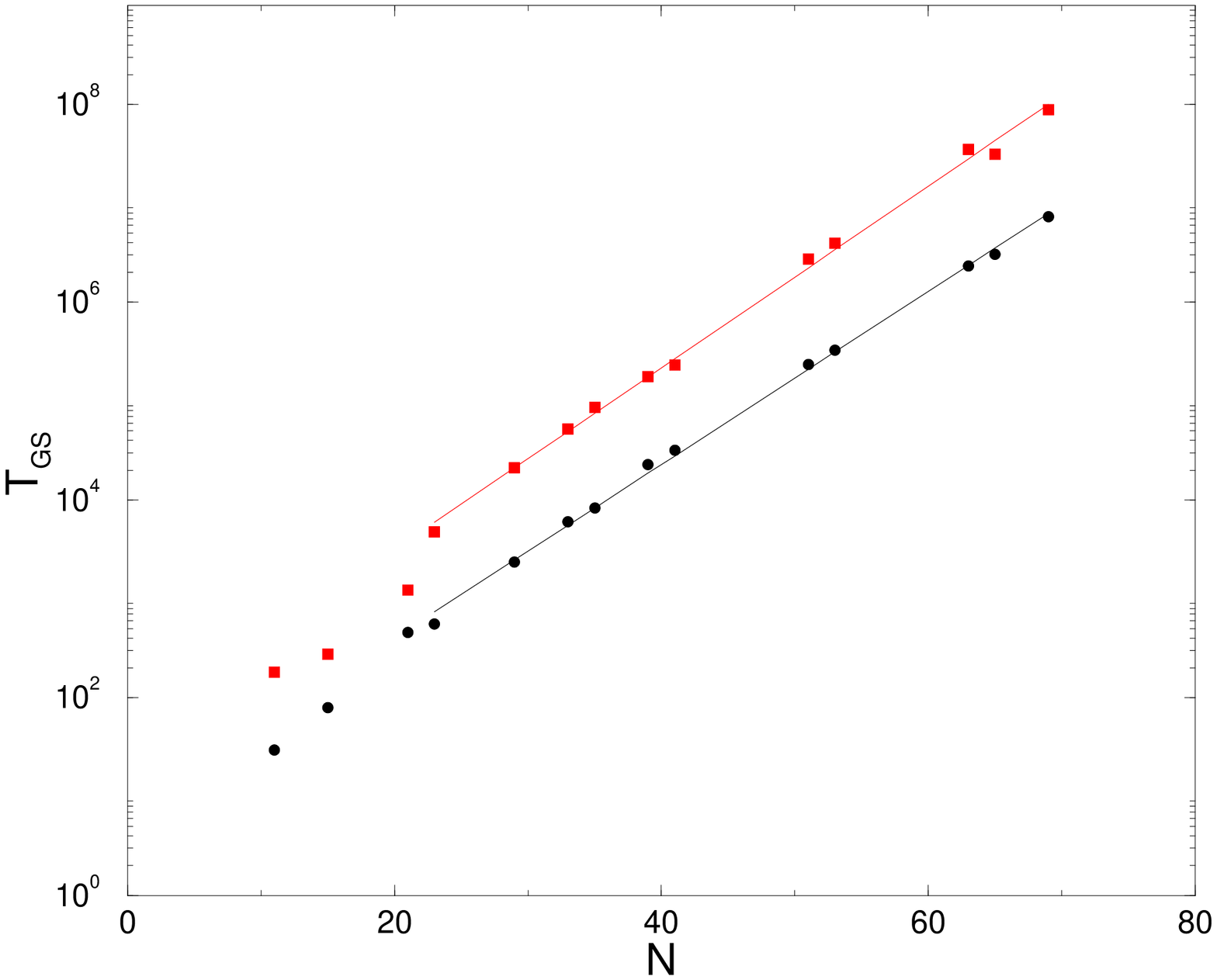}
\caption{
{\small Average time to find the ground state for
greedy (circle) and reluctant (squares) dynamics
for the Sine model.
The straight lines are fits to exponential law
$T_{GS}\sim e^{0.20N}$}
}
\label{seno-total}
\end{center}
\end{figure}

\noindent
On the other hand, if one measures the average
time to find a ground state, Eq. (\ref{av_time_gs}),
which takes into account both the average time
of dynamic and the probability of finding the
ground state,
one can see from Fig (\ref{seno-total}) that greedy
algorithm requires a smaller time on average.
This means that, from a practical point of view,
for a given elapsed time greedy dynamic is slightly more
efficient.

\section{Results for the Sherrington-Kirkpatrick \\
model}

The Sherrington-Kirkpatrick model is the infinite range
case for spin-glasses \cite{SK}. The couplings
$J_{ij}$ are independent identically distributed
symmetric gaussian random variables
$(J_{ij}=J_{ji},\;J_{ii} = 0)$ with
zero mean and variance $1/N$. Since this is a
disordered model one is interested in the
quenched average ground state energy.
For each $N$ this is defined as:
\be
e^{GS}_N = \av{\frac{1}{N}{\inf}_{\s} H_N(J,\s)}
\ee
where we denoted by $\av{}$ the average over
the couplings. Analytical knowledge of this
quantity is available in the thermodynamical limit
$N\rightarrow\infty$ using Parisi Ansatz for
replica symmetry breaking theory:
$e^{GS}_{\infty} = -0.7633$ \cite{MPV},
while numerical simulations obtained using
finite size scaling $e^{GS}_{\infty} = -0.76 \pm .01$
\cite{KS}, $e^{GS}_{\infty} = -0.755 \pm .010$ \cite{TAP},
$e^{GS}_{\infty} = -0.775 \pm .010$ \cite{PP}.

The statistical analysis on the Sine model revealed
that, for a given number of initial conditions,
reluctant dynamic works better than greedy to find the
lower states in energy landscape. On the other hand,
since the reluctant path is much longer than greedy,
from a practical point of view, for a given elapsed time,
it is slightly more efficient to make many quick greedy trials
than a few slow reluctant runs.
For the SK model it is not possible to perform
the same analysis, because a complete control
of the ground state is lacking and also it fluctuates
from sample to sample. To check the conclusion
of previous Section we thus performed a series
of numerical experiments varying control parameters.

\begin{figure}[h]
\begin{center}
\includegraphics[width=10.cm,angle=-90]{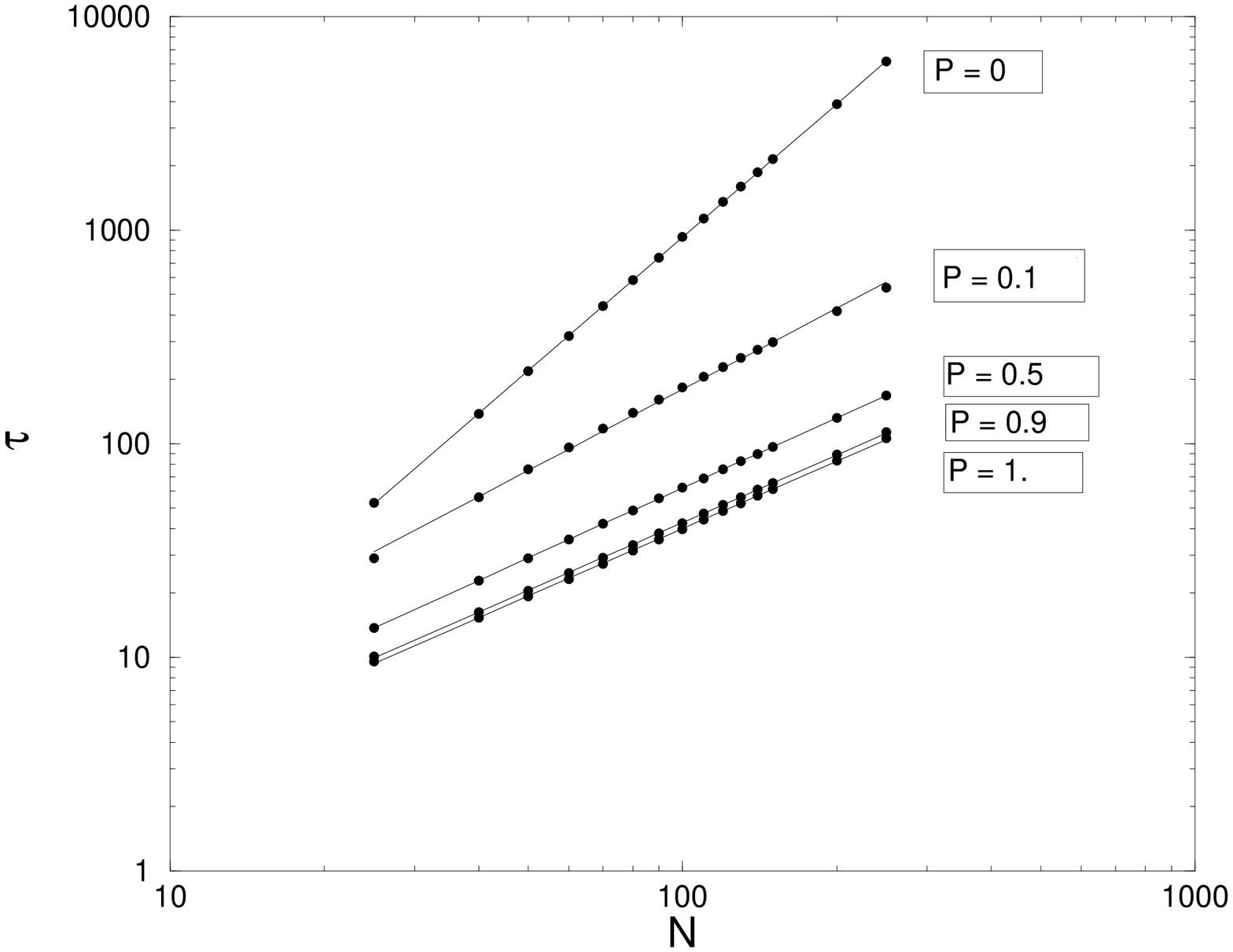}
\caption{
{\small The average time to reach a metastable configuration
for the SK model for different values of $P$. Top to bottom:
$P = 0$ (reluctant), $P=0.1$, $P=0.5$, $P=0.9$, $P=1$ (greedy).
The continuous lines are the numerical fits to power law: $\tau(N) \sim
N^{\alpha}$, with $\alpha = 2.07, 1.26, 1.08, 1.05, 1.04$
from top to bottom.}
}
\label{sk-intrinsic}
\end{center}
\end{figure}

Moreover we investigate the efficiency of a stochastic convex combination
of the two algorithms: with probability $0\leq P \leq 1$ we perform
a greedy move and with probability $1-P$ the corresponding
reluctant move. The deterministic
dynamics are obtained at $P=1$ (greedy) and
$P=0$ (reluctant) respectively. Intermediate
values of $P$ are stochastic dynamics where
the greedy and reluctant moves are weighted
by the probability $P$. First of all we probed
the average time of the dynamic for different
values of $P$ using formula (\ref{time}),
which is easily accessible to measurements
and has good self-averaging properties.
Results are shown in Fig. (\ref{sk-intrinsic}),
together with the best numerical fits.
Note the progressive increase of the slope in log-log scale
from an almost linear law for greedy (bottom)
$\tau_{\{P=1\}}(N)\sim N^{1.04}$
to an almost quadratic law for reluctant (top)
$\tau_{\{P=0\}}(N)\sim N^{2.07}$. However an interesting result is that for
$P=0.1$ we have still have $\tau_{\{P=0.1\}}(N)\sim N^{1.26}$,
i.e. a stochastic algorithm which makes on average
one greedy move (and nine reluctant moves) out of ten
has a much smaller average time than the deterministic
reluctant algorithm $P=0$.
We notice that the exponents for greedy (resp. reluctant) algorithm are
very close to the integers 1 (resp. 2) with an observed slow crossover
between the two for intermediate $p$. It would be interesting to have
a theoretical understanding of this fenomenon even if only at a
heuristic level. We plan to return over this problem in a future work.

\begin{figure}[h]
\begin{center}
\includegraphics[width=10.cm,angle=-90]{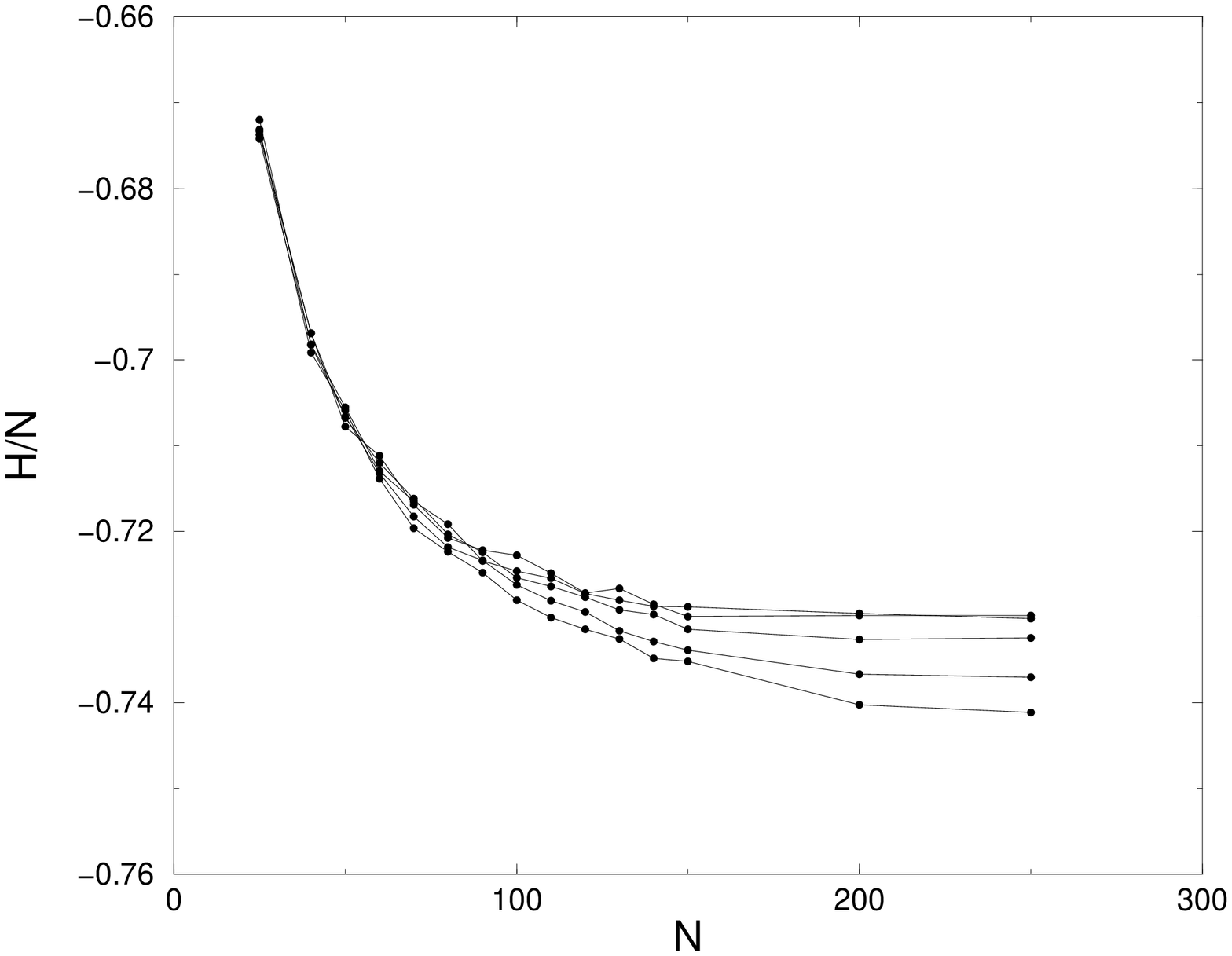}
\caption{
{\small Lowest energy value using a protocol of choosing
$N$ initial condition for the SK model for different value
of $P$. Bottom to top:
$P = 0$ (reluctant), $P=0.1$, $P=0.5$, $P=0.9$, $P=1$ (greedy)}
}
\label{sk-fix-init-cond}
\end{center}
\end{figure}

\noindent
Next we measured the lowest energy value found
for a given number of initial conditions for
different probability $P$. One has to choose a
protocol to fix the number of initial conditions.
Obviously, the larger the system size the bigger
must be the number of trials.
We tried different choices obtaining similar results.
For the sake of space we show in Fig. (\ref{sk-fix-init-cond})
the results of the run where we choose $N$ initial
conditions for a system of size $N$. The data
have been averaged on $1000$ disorder realizations.
We see that the smaller is the probability of
making greedy moves, the lower is the energy
found. The best results is obtained for $P=0$,
which corresponds to deterministic reluctant
dynamic. This confirms that, ignoring the total
amount of time and imposing constrain only
on the number of initial conditions,
reluctant dynamic is the most efficient
in reaching low energy states.

\begin{figure}[h]
\begin{center}
\includegraphics[width=10.cm,angle=-90]{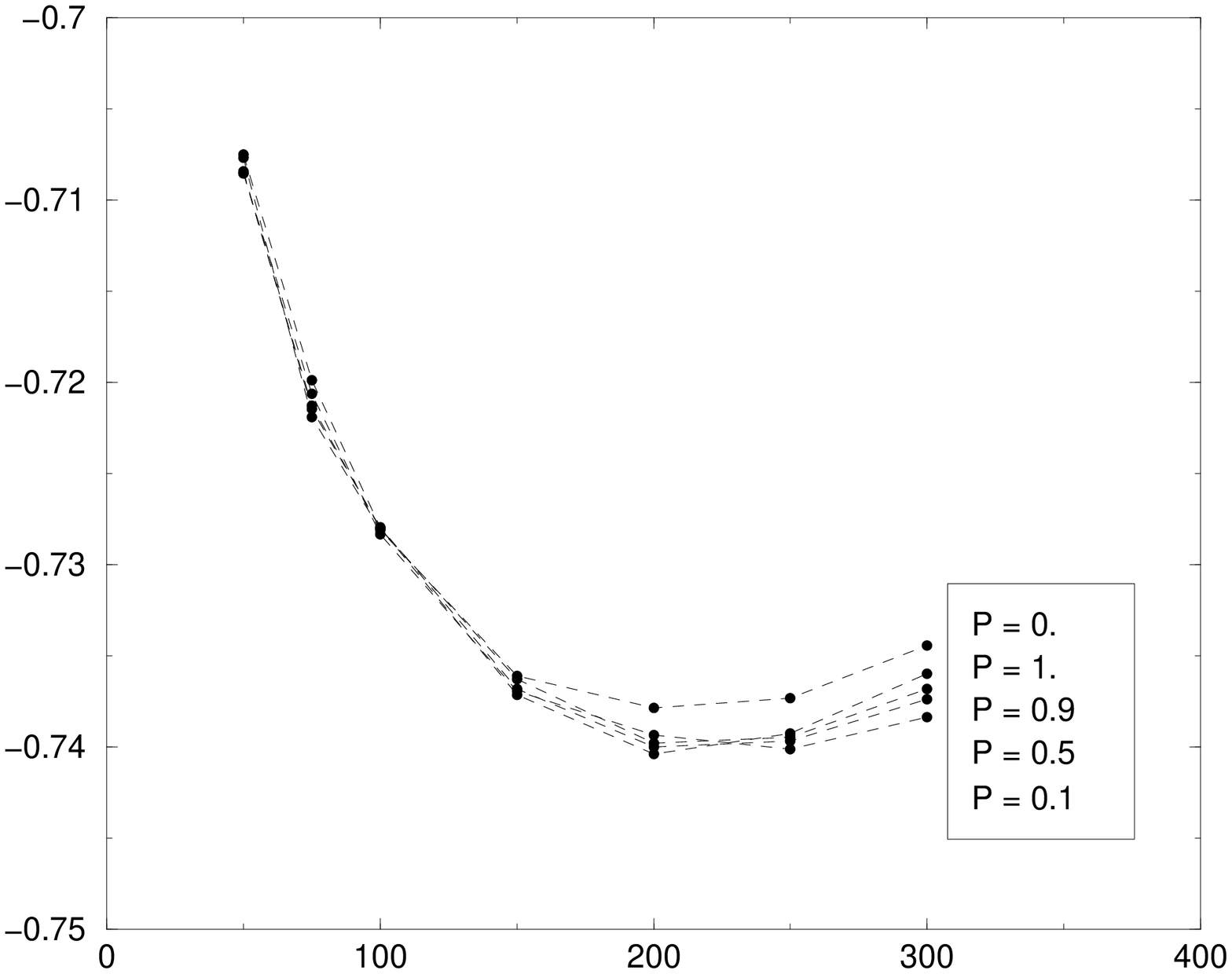}
\caption{
{\small Lowest energy value for a fixed elapsed time
of $100$ hours on a CRAY SP3 for the SK model for
different value of $P$ (see legend).}
}
\label{sk-fix-elapsed}
\end{center}
\end{figure}

\noindent
Finally we compared results of different
probabilities in the case one considers
a fixed elapsed time. As an example,
we present results for an elapsed time
of $100$ hours of CPU on a CRAY SP3 for
$N$ in the range $[50,300]$.
We considered again $1000$ disorder realizations
and assigned the same time length to each sample
($6$ minutes). Obviously in this way
reluctant dynamic starts from a smaller
number of initial conditions than greedy,
because its relaxation time is longer.
In Fig. \ref{sk-fix-elapsed}
we plot the values of the lowest energy state
as a function of $N$.
We can see from the data that,
for a fixed elapsed time, greedy
dynamic ($P=1$) find lower energy
states than reluctant ($P=0$).
Moreover we observe that the best result
is obtained for $P=0.1$. Thus we suggest
that the more power full strategy to find
low energy state using greedy and reluctant
dynamic is a combination of them, where
most of the steps the move is reluctant
and on a small fraction of steps (say $0.1$)
the move is greedy.

\vspace{0.5 cm}
\noindent
{\bf Acknowledgments}

\noindent
We would like to thank CINECA for the computation grant
on the CRAY SP3 and  Prof. I. Galligani for his encouragement
and his support. We thank the referees for their suggestions 
and comments.

\end{document}